\documentclass[prb,preprint,amsmath,amssymb]{revtex4}
\usepackage{color}
\begin{document}
\author{Kevin Leung$^*$}
\affiliation{Sandia National Laboratories, MS 1415, Albuquerque, NM 87185\\
$^*${\tt kleung@sandia.gov}}
\date{\today}
\title{DFT Modelling of Explicit Solid-Solid Interfaces in Batteries:
Methods and Challenges}

\input epsf.sty
 
\begin{abstract}

Density Functional Theory (DFT) calculations of electrode material properties
in high energy density storage devices like lithium batteries have been
standard practice for decades.  In contrast, DFT modelling of explicit
interfaces in batteries arguably lacks universally adopted methodology 
and needs further conceptual development.  In this paper, we focus on
solid-solid interfaces, which are ubiquitous not just in all-solid state
batteries; liquid-electrolyte-based batteries often rely on thin, solid
passivating films on electrode surfaces to function.  We use metal anode
calculations to illustrate that explicit interface models are critical
for elucidating contact potentials, electric fields at interfaces, and 
kinetic stability with respect to parasitic reactions.  The examples emphasize
three key challenges: (1) the ``dirty'' nature of most battery electrode
surfaces; (2) voltage calibration and control; and (3) the fact that
interfacial structures are governed by kinetics, not thermodynamics.  To meet
these challenges, developing new computational techniques and importing
insights from other electrochemical disciplines will be beneficial.

\end{abstract}
 
\maketitle
 
\section{Introduction}
\label{introduction}

High energy density batteries are instrumental to vehicle electrification
and to grid storage which helps alleviate intermittency in solar and
wind energy generation.  Deeper understanding of existing materials and
the search of new electrode and electrolyte materials are crucial
for developing higher capacity, faster-charging, safer, and longer-lasting
batteries.\cite{whittingham}  Battery interfaces are also
universally acknowledged to be critical for governing battery rate 
capability and stability.\cite{bes}  Charge/discharge processes, as well
as many side reactions that lead to degradation, self-discharge, and
thermal runaway, initiate at interfaces.

Electronic structure Density Functional Theory (DFT) modelling of the
crystalline interior of electrodes, for the purpose of predictng phase
stability, equilibrium voltages, and lithium diffusion kinetics, has been
widely practised for decades.\cite{ceder,ceder1,islam1,islam2,islam3}  It
provides important guidance to experiments.  In contrast, DFT modelling of
explicit battery interfaces arguably needs further conceptual development and
systematization.  Here we distinguish explicit interfaces where two phases are
in contact in the same simulation cell, from single-phase DFT calculations used
to infer interfacial properties indirectly.\cite{phase} The present work
focuses on modelling methods, offers somewhat pedagogical discussions on the
rationale behind the DFT approaches used in our group, and examines future
directions and improvements.  Simple material interfaces, mainly involving
lithium metal, are used as illustrations, but the principles involved should
be broadly applicable to other electrodes.  Even simple materials
exhibit complex interfaces that require substantial approximations; the nature
and consequences of some key implicit approximations are highlighted.  We
restrict ourselves to vanishing current densities.\cite{prendergast} This
paper is intended to be a topical, critical overview, not a comprehensive
review of battery interface DFT modelling.\cite{qi_review,borodin17,butler}

\begin{figure}
\centerline{\hbox{ \epsfxsize=2.50in \epsfbox{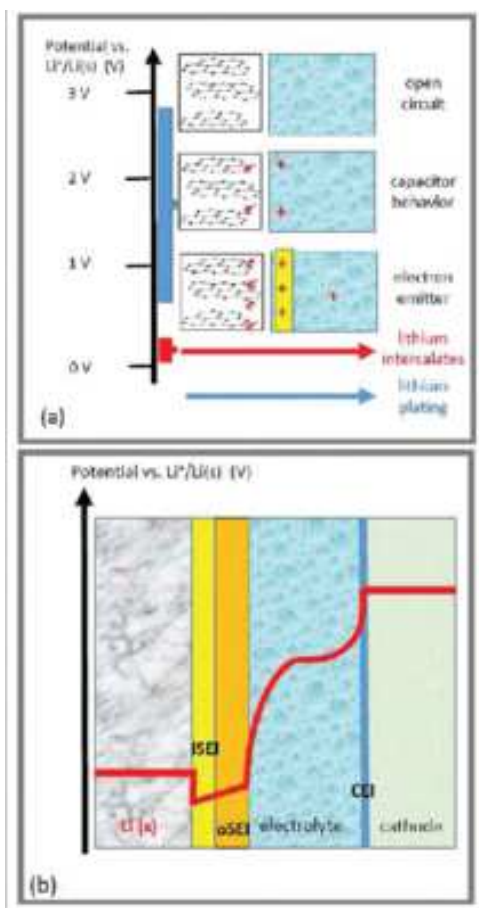} }}
\caption[]
{\label{fig1} \noindent
Schematics illustrating
(a) first charge of a graphite anode in lithium ion batteries;
(b) proposed voltage profile on battery with ``dirty'' electrode surfaces.
``iSEI'' and ``oSEI'' refer to inorganic and organic SEI components.
The iSEI has nanometer thickness.  The separator occupies the electrolyte
region and is omitted.
}
\end{figure}

A useful illustration of the multiplicity of interfacial processes is the
first charge of graphite anodes used in commercial liquid electrolyte-based
lithium ion batteries (Fig.~\ref{fig1}a).  The ``open circuit'' potential
(OCP) of pristine graphite in organic carbonates, e.g., ethylene carbonate,
EC, mixed with salts and cosolvents, is $\sim$3~V vs.~Li$^+$/Li(s)
(henceforth this is the reference used unless noted).  We stress that OCP
is measured by immersing the electrode in the electrolyte over experimental
time scale, impling that the interface is well-equilibrated.  As charging
begins and the voltage drops, graphite acts as an inert electrode like
those in electrochemical capacitors.  Excess $e^-$ go to the surface and
their negative charges are compensated by an enhanced local concentration
of cations in the electrolyte near the interface; the familiar electric double
layer (EDL, effectively a dipole sheet) is formed.  As the voltage approach
$\sim$0.7~V, graphite acts as an electron emitter --- possibly also as an
electrocatalyst at its edge site functional groups --- and the organic solvent
is electrochemically reduced in what will be called ``parasitic reactions.''
This is because high energy batteries typically operate at voltages outside
the electrochemical stability limit of electrolytes.\cite{xu2004,stability,batt}
(Counterions can be reduced at higher voltages, more slowly and to a lesser
extent.\cite{borodin11,borodin15,vedge}) The passivating ``solid electrolyte
interphase'' (SEI)\cite{xu2004,qi_review,aurbach_mg} grows from these
decomposed electrolyte fragments, but $e^-$ can tunnel or diffuse through the
nascent film, possibly as part of Li atoms, until later charging cycles.  The
EDL, which sustains the applied voltage, may now partly reside in the solid
film, not just in the liquid (Fig.~\ref{fig1}b).  Li$^+$ does not intercalate
into graphite sheets until 0.2-0.1~V (not counting edge sites).  This small
0.1~V voltage window has seen the majority of DFT studies,\cite{persson}
in the bulk Li$_x$C$_6$ region.  By continuity, in this regime, the voltage
must be a manifestation of both the graphite electronic structure which changes
with $x$, and the EDL contribution.  The SEI continues to grow/evolve in
this window.  When a LiC$_6$ stoichiometry is attained, the fully charged
graphite reverts to an inert electrode, albeit coated with SEI, that may
plate lithium metal at negative voltages.  Lithium plating is detrimental to
battery life and safety.  Related processes occur on transition metal oxide
cathode surfaces, on which ``cathode electrolyte interphase'' (CEI) is
formed (Fig.~\ref{fig1}b); however, the voltage-dependent parasitic reactions
there most likely involve species in contact with the surface, just like in
water splitting, rather than long-range electron transfer like on the anode.

Thus, as our previous overview emphasized,\cite{interface1} batteries
interfaces embody the rich physics of mulitple electrochemical devices, and
DFT-based battery modelling benefits from borrowing from diverse computational
disciplines.  Electrified liquid/solid interfaces on passive, pristine
carbon or platinum surfaces are relevant to electrochemical capacitors
and electrocatalysis; these areas have well-established DFT modelling
protocols.\cite{otani,otani12,otani13,mira,mira2014,neurock,sugino1,arias,dabo1,marzari,sprik}
Transition metal oxide/liquid interfaces in batteries are in some ways 
related to electrochemical and photoelectrochemical water
splitting.\cite{sprik_tio2,sprik_tio2a,selloni_tio2,galli_tio2}
Electrodes covered with passivating solid films are relevant to breakdown
of surface oxide films which protect metal surfaces against corrosion,
and constructions similar to Fig.~\ref{fig1}b have been invoked in
corrosion studies.\cite{macdonald,marks,costa}  Electroplating studies
are also of obvious interest to batteries.\cite{schmickler}

Passivating films on organic-solvent-based battery anode surfaces have
inorganic (iSEI) and organic (oSEI) 
components (Fig.~\ref{fig1}b),\cite{aurbach_two_layer} leading to multiple
solid-solid interfaces.  We argue that it is as urgent to model these
comparatively neglected inorganic solid interfaces as the solid-liquid
interfaces which have been the mainstay of computational electrochemistry.
DFT simulations have been conducted on explicit battery solid-solid interfaces
relatively recently,\cite{santosh,holzwarth,solid,gb,batt,ucsd,sodeyama,tateyama1,tateyama2,sumita}
partly because all-solid-state batteries with ceramic- or sulfide-based solid
electrolytes constitute a timely research area.  Many aspects of solid-solid
interface studies in liquid-electrolyte batteries are informed by and are
transferrable to these solid electrolytes.\cite{batt,gb} The main difference
is the solid film/electrolyte thickness --- nanometers vs.~microns.  The
much thicker solid electrolytes come with a wider ``space charge'' 
region\cite{voltage_map,kpfm} which can potentially be modelled using
continuum methods coupled to DFT currently applied to the Mott-Schottky layer
at semiconductor interfaces.\cite{dabo1}
One theme common to both is that the charge carrying cation M$^+$ (Li$^+$ and
Na$^+$) is highly mobile by solid state conductivity standards (otherwise
the material would not be used in batteries).  Hence the total number of
M~atoms should vary with voltage at the interface as well as inside electrodes.

We identify three computational challenges somewhat unique to battery
interfaces.  One is the ubiquity of the aforementioned solid-solid
interfaces.  Solid surface films in which M$^+$ can diffuse limit the
utility of modelling techniques traditionally used in the liquid state like
DFT-based molecular dynamics (also called {\it ab initio} molecular dynamics
or AIMD).\cite{qi_review,nasa,goddard,bal_lioh,budzien}  This is because of
time scale mismatch: ionic motion and relaxation in solid 
occur many orders of magnitude slower than liquid state diffusion.  

Another critical issue is voltage determination and control.  First we stress
that an assumption often invoked when modelling explicit battery interfaces,
namely that voltages in interfacial simulation cells are determined by the
lithium content/energetics, is incorrect.  Instead,
we adopt the definition used in other electrochemical
disciplines\cite{otani,otani12,otani13,mira,mira2014,arias,dabo1,marzari,voltage,sprik} 
and reconcile it with traditional battery modelling approaches.  Using  the
correct definitions is necessary to estabilish equilibrium, and to construct
simulation cells at overpotentials conditions.

\begin{figure}
\centerline{\hbox{ \epsfxsize=5.0in \epsfbox{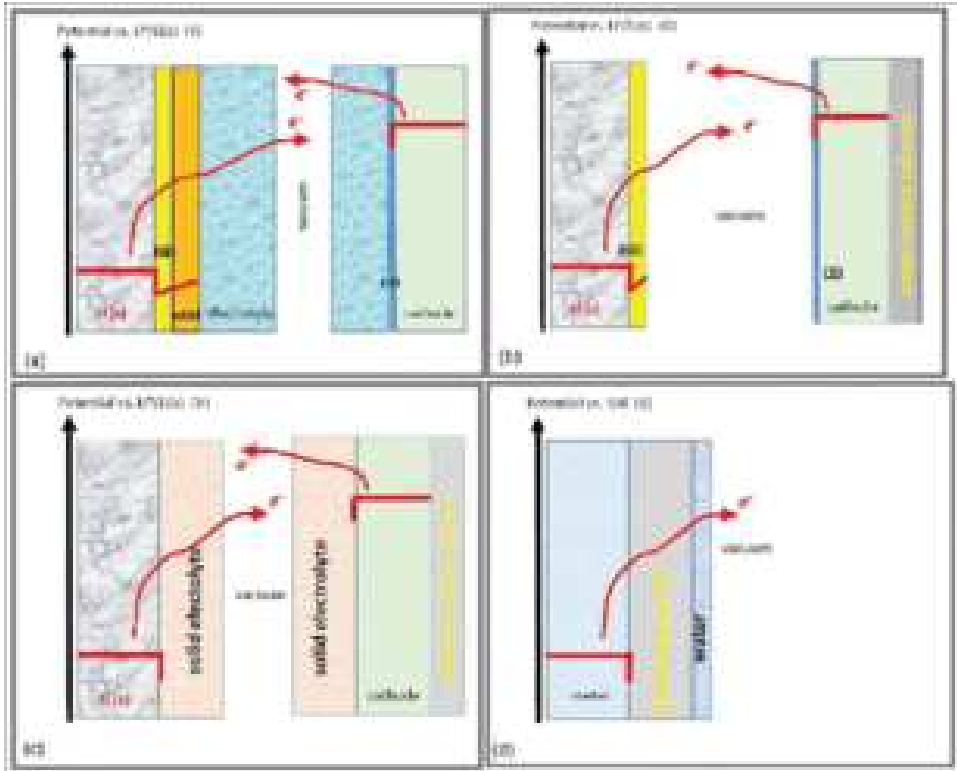} }}
\caption[]
{\label{fig2} \noindent
(a) Inserting a vacuum region into the liquid electrolyte of Fig.~\ref{fig1}b.
This step is rigorous.
(b) Reduced anode model and cathode models, introducing approximations.
A metallic current collector is added at the backside of the cathode, 
(c) Schematic of solid electrolytes with vacuum inserted.
(d) Schematic of metal oxide-coated metal surface, relevant to corrosion.
Arrows with electrons indicate formal paths for calculating ${\cal V}_e$;
in physical systems it may be ions that cross the boundaries.  M=lithium in
this figure.
}
\end{figure}

Finally, battery interfaces are determined by kinetics, not thermodynamics.
Inorganic electrode and electrolyte materials are often sintered for hours
at $\ge$900~$^o$C.  A phase diagram, thermodynamics modelling approach suffices
under those conditions.\cite{phase}  In constrast, interfaces that involve
organic solvents are assembled at room temperatures where chemical reactions
are slow.  Even all-solid-state battery interfaces are usually created
at modest ($<$200~$^o$C) temperatures, where thermodynamic equilibrium does
not always apply,\cite{lipon} although thermodynamic phase
diagrams for bulk\cite{phase} and surfaces\cite{mira2018b} provide the
foundation for kinetics investigations.  Thermodynamic (single-phase)
calculations are elegant and systematic.\cite{butler,phase}  Interfacial
kinetics calculations tend to be idiosyncratic.  They depend on the realism
of the model interface, lattice matching, and other issues common to all
interface modelling efforts.  Here we focus on issues unique to
electrochemical interfaces.

It must be stressed that a reaction with a large exothermicity
($\Delta G$) does not necessarily exhibit a low barrier ($\Delta G^*$).  One
famous example is the Grotthuss mechanism for proton exchange in water:
\begin{displaymath}
{\rm H}_2{\rm O} + {\rm OH}^- \rightarrow {\rm OH}^- + {\rm H}_2{\rm O} . 
\end{displaymath}
The left and right side are the same, and $\Delta G$ is exactly zero.  Thus
the thermodynamic driving force is the lowest possible; a more unfavorable
$\Delta G$ would stop the reaction from occurring.  Yet the reaction is
almost barrierless ($\Delta G^*$=0).\cite{tuckerman}  Hence kinetics
and thermodynamics are not always correlated.  

Kinetics is particularly relevant to the viability of conversion cathode materials\cite{wolverton} and to stability against parasitic
reactions.\cite{qi_review} Currently DFT modelling of conversion cathodes is
focused on elucidating metastable intermediate states,\cite{wolverton} not on
computing voltage-dependent reaction barriers;\cite{chan} in electrocatalysis,
both intermediates and barriers are known contribute to
overpotentials.\cite{chan}   Finally, kinetics approaches may not identify the
final products.  To circumvent this, whenever possible, we have focused on
elucidating the stability of proposed surface film components rather than
start parasitic reaction calculations from pristine electrode
surfaces.\cite{batt}

\section{Methods and Models}

DFT calculations only deal with half cells because DFT is a ground state
theory which supports a single Fermi level ($E_{\rm F}$).  Hence we insert
a vacuum region into Fig.~\ref{fig1}b, generating two half cells
(Fig.~\ref{fig2}a).  This construction is rigorous for liquid electrolyte
systems\cite{trasatti} and has been applied in DFT modelling
work.\cite{neurock,sugino1}

\subsection{Models of Thin-Film-Covered Anode Interfaces}
\label{models}

Focusing on the anode, the multilayer SEI model\cite{aurbach_two_layer}
suggests that inorganic products coat the active anode material, separating
it from the thicker, amorphous, and porous organic/polymeric layer outside.
The organic SEI is complex, and likely exhibits a low dielectric
constant.  Its micro- and even atomic-structures are subjects of current
studies and debate.\cite{luning,villanova,utaustin} SEI chemical compositions
also evolve with time.\cite{evolve1,evolve2} 

To make progress, we replace the outer organic SEI (oSEI), with vacuum
(Fig.~\ref{fig2}b).  This model focuses on the anode active material/innermost
inorganic SEI (iSEI) interface, expected to play a main role in blockage of
$e^-$ tunneling to the electrolyte.  Our liquid-free,
charge-neutral simulation cells resemble those we use to model all-solid-state
battery interfaces (Fig.~\ref{fig2}c).\cite{lipon}  In principle, the omitted
electrostatic contribution from a liquid electrolyte can be computed using
classical force-field-based molecular dynamics, but this is not applicable when
an organic SEI covers the inorganic film.  (To give an order-of-magnitude
estimate, the liquid EC/vacuum interface exhibits a $\sim$0.3~V surface
potential,\cite{voltage} dwarfed in magnitude by solid-solid interface effects
discussed below.) Another approach might be to replace the vacuum in
Fig.~\ref{fig2}b with implicit solvation.\cite{otani,arias,austrian}
While noble metal-liquid interactions have been calibrated in
implicit solvation models, oxide/water or oxide/organic liquid interactions
have not.  Currently no dielectric solvation model is universally
available or uniformally implemented into every DFT code.  To enhance
reproducibility, we work with vacuum interfaces.  The EDL resides entirely
within the thin solid film regions in our models.  Our approach is clearly
a limiting-case approximation, to be improved upon in the future.  

Our half-cell model for solid-electrolyte/electrode interfaces
(Fig.~\ref{fig2}c) has an unavoidable ambiguity: the absolute work function
$\Phi$ (and hence the estimated voltage, see below) depends on the nature of
the cleaved facet, unlike liquids which relax to universal liquid-vapor
interfaces.  Kelvin Probe Force Microscopy (KPFM) measurements of
voltages in solids share this ambiguity.\cite{kpfm}  The voltage difference
between cathode and anode does not depend on the cleaved surface provided the
two vacuum-solid electrolyte interfaces are identical.  Vacuum regions
in fact exist in solid electrolytes, inside pores and cracks.\cite{pores}

\subsection{Voltage Definitions for Metallic Electrodes}
\label{voltage}

Unique among electrochemical devices, battery electrodes can emit both
electrons and the charge-carrying cation M$^+$.
The primary battery functions are charge and discharge, which involve M$^+$
intercalation and exit from electrodes into the electrolyte, accompanied
by $e^-$ flow in the external circuit.  However, as discussed above, $e^-$
can also be emitted into (reduction) or captured from (oxidation) the
electrolyte, accompanied by local M$^+$ redistribution but not M$^+$ transfer
between electrodes.  

To account for these separate functions, we define two voltages, ${\cal V}_i$
and ${\cal V}_e$.\cite{voltage}  The ``ionic'' voltage ${\cal V}_i$ is the
canonical definition in the battery DFT literature.\cite{ceder} It
measures the energy gain by inserting an M~atom, referenced to
M(s)$\rightarrow$M(atom):
\begin{equation}
{\cal V}_i=[(E_{n_{\rm M}}-E_{n'_{\rm M}})/(n_{\rm M}-n'_{\rm M})
	- \mu_{\rm M}]/|e|,
							\label{eq1}
\end{equation}
where $E_{n_{\rm M}}$ is the total energy of simulation cell with an
electrode with $n_{\rm M}$ M atoms, $\mu_{\rm M}$ is the M chemical
potential in its bulk metal phase, and $|e|$ is the electronic charge.
The entropy change is small in solids and is generally neglected in
Eq.~\ref{eq1}.  ${\cal V}_e$ is the widely used definition of voltages
in computational electrochemistry; it is
$\Phi$=$(E_{\rm vacuum}$-$E_{\rm F}$)/$|e|$ modulo a constant, where
$E_{\rm vacuum}$ is the vacuum level in the same simulation cell that
there Fermi level $E_{\rm F}$ is computed.  $e^-$ is the prime mover in
electrochemistry; potentiostats and voltmeters control and measure $e^-$
energy ($E_{\rm F}$), not M~content.  

${\cal V}_e$ is notoriously difficult to define and control in DFT
calculations.  For liquid electrolytes associated with corrosion, fuel
cells, and batteries, AIMD-based thermodynamic integration approaches
have been applied to calibrate the voltage in condensed phase simulation
cells without vacuum regions.\cite{otani,mira,mira2014,voltage,vedge} 
(At least some of these works\cite{voltage,vedge} were influenced by the
pioneering research of Sprik and coworkers.\cite{sprik2012,sprik2008})
Such methods calculate half-cell electrochemical potentials via free
energy changes; the predictions are solvent-dependent.  

The majority of ${\cal V}_e$ calculations rely on having a vacuum region (or a
quasi-vacuum containing an implicit solvent) in the simulation cell.  Vacuum
is in effect the reference electrode.  In periodic boundary DFT calculations,
both finite temperature AIMD\cite{otani12} and T=0~K configuration optimization
calculations\cite{neurock,arias} have used vacuum to compute absolute
voltages.  The Berry's phase method, under development for liquid
electrolytes,\cite{sprik} can potentially circumvent the need for vacuum.

The relation between voltage referenced to Standard Hydrogen Electrode
(SHE), and the work function ($\Phi$), was famously
quantified by Trassati for liquid electrolytes.\cite{trasatti}  It is
$\Phi$ minus 4.44~eV divided by $|e|$.  This relation has been incorporated
into or used as calibration for implicit solvent formulations of voltage
calculations.\cite{otani,marzari,arias} In cluster-based quantum chemistry
and DFT calculations, where only a solute and sometimes a few solvent
molecules are surrounded with implicit solvent, calibration is not possible,
and the Trasatti relation is directly applied to relate electron affinities
(EA) and ionization potentials (IP) to redox potentials.\cite{homo,borodin11}
When the SHE reference is replaced with Li$^+$/Li(s), the shift is changed
from 4.44~eV to between 1.37 and 1.40 eV.  The O(0.01 eV) differences are
within DFT uncertainties.  

A metallic electrode is just a large molecule where EA=IP, with the exception
that different facets may have different $\Phi$'s which are reconciled by
charge transfer and/or geometric considerations.\cite{mermin}  The complex
organic SEI (Fig.~\ref{fig2}a) precludes calibration calculations here.  Hence
in this and our previous work, we simply assign
\begin{equation}
{\cal V}_e = \Phi/|e| -1.37~V, \label{eq2}
\end{equation}
to electrodes in our overall charge-neutral simulation cells, just like for
small molecules, and acknowledge the approximation in omitting the material
outside the inorganic solid.  Trasatti likely never intended the use of
Eq.~\ref{eq2} in vacuum, but one can substitute a low dielectric electrolyte,
like argon at low density.  We also use this relation for pitting
corrosion calculations, where water is optimized at T=0~K.  In the future,
we will switch to sampling H$_2$O at T=300~K, which is the more rigorous
approach.  Unlike Ref.~\onlinecite{neurock}, which also opens a
vacuum gap in the frozen bulk liquid electrolyte region, we include at most
a sub-monolayer of water molecules.  Like Refs.~\onlinecite{mira,voltage,vedge}
and unlike Refs.~\onlinecite{otani,arias}, we control ${\cal V}_e$ using atoms
and $e^-$, not with effective medium approaches extrinsic to standard DFT.

\subsection{$i$-Equilibrium}

We define ``$i$-equilibrium'' to be established when ${\cal V}_e$=${\cal V}_i$.
Rigorously speaking, $i$-equilibrium is required for comparison with
constant-voltage condition experiments.  The exact $i$-equilibrium condition
is affected by the accuracy of the DFT functional used, which is neglected
herein.  Since the two electrochemical potentials follow the relation
${\bar \mu}_{\rm M}$=${\bar \mu}_{e^-}$ + ${\bar \mu}_{{\rm M}^+}$ 
${\cal V}_e$=${\cal V}_i$ implies a specific ${\bar \mu}_{{\rm M}^+}$ 
in the electrolyte\cite{voltage} which can be used for future calibration
calculations.  When $i$-equilibrium is violated, the system is at an
overpotential with respect to
at least one process.  Overpotentials often occur in batteries because
M$^+$ and $e^-$ motions are generally separated by orders of magnitude in time
scales.  Experimentally, M$^+$=Li$^+$ relaxation times in cathode materials can
be minutes.\cite{gitt} For this reason, we also call ${\cal V}_e$ and
${\cal V}_i$ the ``instantaneous'' and ``equilbrium'' voltages, respectively.
In DFT calculations, Li$^+$ relaxation and motion are far more limited than 
in experiments because the time scale (if using AIMD) is much shorter,
and there is no Li reservoir.  Therefore DFT modelling of interfaces is
even more susceptable to overpotentials.  As long as the ``electrode'' (a large
clump of atoms) is metallic, ${\cal V}_e$ is instantaneously well defined, even
if there are forces on atoms.  ${\cal V}_i$ is in contrast only well-defined
when atomic forces are zero at zero temperature, or when the Boltzmann
distribution is satisfied at finite $T$.

An incorrect assumption often made\cite{ucsd} (including in our early
work\cite{budzien}) is that simulation cells containing an interface is
automatically at ``open-circuit'' voltage equal to ${\cal V}_i$.  This ignores
the possibility that $i$-equilibrium can be violated.  In fact, ${\cal V}_e$
is seldom discussed or reported when modelling batteries.  The reason is likely
historical.  The majority of DFT calculations on electrode materials deal with
the crystal interior (i.e., they are ``single phase'').  In the absence of
interfaces, the average electrostatic potential in the simulation cell is
undefined to a constant\cite{dutch} except in special cases.  Hence the
absolute $E_{\rm F}$, which depends the average potential, is also underfined
--- as is ${\cal V}_e$ via Eq.~\ref{eq2}.  The only logical recourse then
is to assume $i$-equilibrium.  When an electrode/electrolyte
interface exists in the simulation cell, however, a potential drop occurs that
can in principle be computed (Fig.~\ref{fig2}).  Other reasons interfaces
may be out of equilibrium include the presence of high current
densities, and electrode polarization whereby a liquid electrolyte responds
too slowly to voltage changes.  These are not addressed herein.

\subsection{Grand Canonical Ensemble for $e^-$ and Li$^+$}

Excess $e^-$ surface densities on pristine electrode surfaces vary during
charge and discharge of EDL capacitors, fuel cell electrodes, and other
devices.  As a result, mobile ions redistribute at interfaces and EDL's
are modified.  Grand canonical treatments of $e^-$, which permits fractional
$e^-$ compensated by an effective medium, have been developed for this
purpose.\cite{otani,marzari}

In batteries, charge carriers M$^+$ are also mobile; their
concentrations evolve as voltage varies not only inside battery anodes and
cathodes, but also at interfaces where there is often less crystalline order.
To model changing M$^+$ content calls for Grand Canonical Monte
Carlo (GCMC) or related methods, in conjunction with electronic structure
calculations.  AIMD simulations, one of the mainstays in battery interfacial
calculations, lack this capability.  MC is time-independent and can in
principle circumvent the orders-of-magnitude solid-liquid time scale mismatch.
In the past, DFT/MC calculations were costly because the global DFT energy
was recomputed after each single-atom motion in a MC trial.\cite{siepmann}
GCMC/DFT-like calculations are being implemented;\cite{rappe,sugino} they are
extremely promising for future studies of battery interfaces.  Until such
methods become widely available, the interfacial M$^+$ content has to be
varied manually or via high-throughput calculations, as it has been done
in single phase battery electrode calculations.\cite{ceder,ceder1}

Many battery cathode materials are polaron-conducting transition metal oxides
with finite band gaps.  Rigorously speaking, ${\cal V}_e$ is undefined such
materials (unless surface/interface regions become metallic\cite{liverpool});
$E_{\rm F}$ is pinned by defect/dopants levels not typically included in the
simulation cell, and it cannot be said that all regions of the electrode are
at the same voltage.  Constant voltage methods which allow fractional $e^-$
in DFT calculations\cite{otani,marzari} cannot be used because
fractional occupancy of transition metal $d$- or $f$-orbitals is unphysical.
Here one can look to experiments for guidance.  In practical batteries, the
cathode is a composite with active oxides mixed with conductive components.
By adding a metallic slab like lattice-matched Au(001) to the bottom of
spinel Li$_x$Mn$_2$O$_4$, one can recover metal-like $E_{\rm F}$ behavior
(Fig.~\ref{fig2}a).\cite{solid} Regarding the structure of cathode surface
films, much less is known about thinner, hard-to-characterize CEI on battery
cathode surfaces in liquid-electrolyte batteries.  Further discussions
of cathode interfaces will be left to future overviews.

\color{black}
\subsection{Modelling Kinetics and AIMD Simulations}
\color{black}

At least three general approaches have been applied to model interfacial
kinetics in batteries: (1) unconstrained AIMD simulations at or near room
temperature; (2) high temperature AIMD; (3) barrier predictions.
\color{black}
(1) and (2) are both unconstrained and only differ in intention; the former
is meant to directly mimic near-experimental and -device conditions while the
latter is explicitly understood to artificially accelerate timescales.
\color{black}
Because of the computational expense, current AIMD simulations seldom exceed
1~ns in trajectory lengths.\cite{goddard}  So unconstrained AIMD
trajectories\cite{goddard,bal_lioh,budzien} are at least 10$^{12}$ times
shorter than experimental time scales.  M$^+$-motion and bond-breaking
reactions revealed in such AIMD trajectories are the initial steps.
Even if AIMD predicts that the electrolyte is atomized,\cite{goddard}
nucleation of these products to SEI solid phases takes more time.  Logically,
if a change is observed in an AIMD trajectory, it provides important
insights about the mechanism.  If no change occurs, it is possible AIMD
trajectories are just not long enough.

Alternatively, parasitic reactions at interfaces have been examined using
AIMD at substantially elevated 
temperatures.\cite{nasa}  This accelerates reaction rates, but extrapolating
predictions to target conditions (usually room temperature) is non-trivial.
This approach can be extremely useful for discovering mechanisms that are
subsequently quantified via barrier-finding calculations.

Reaction barrier calculations like umbrella sampling and metadynamics
bypass trajectory length limits at the cost of having to choose
pathways.\cite{meta,tateyama,benedek,vedge} In purely solid state
calculations, zero-temperature nudged-elastic-band (NEB) calculations have
become standard.\cite{neb}  From barrier calculations, we estimate mean
reaction rates using the standard transition state theory expression
\begin{equation}
1/t_{\rm ave}  = k_o \exp (-\Delta E^*/k_{\rm B}T), \label{eq0}
\end{equation}
where $\Delta E^*$ is the activation energy and $k_{\rm B}T$ is the thermal
energy at room temperature.  At finite temperatures, free energy barriers
($\Delta G^*$) should be used in Eq.~\ref{eq0}.  We adopt
$k_o$=10$^{12}$/s, within a factor of ten of prefactors used in the literature.
Any step in a proposed multi-step reaction mechanisms is considered
viable if it is exothermic ($\Delta E$$<$0) and if $t_{\rm ave}$ is less
than one hour -- which translates into $\Delta E^*$$<$$\sim$1~eV.\cite{batt}  

Rates may depend on voltages, while voltages are modified by atomic motion in
finite sized simulation cells.  This is because the work function $\Phi$
is altered by the effective dipole moment via
\begin{equation}
\Delta \Phi = 4 \pi \sigma d . \label{eq3}
\end{equation}
$\sigma$ is the surface density of a uniform point dipole sheet, $d$ is the
average dipole magnitude and includes screening effects, and atomic units
are used.  This is an ``interface'' effect but the $\Delta \Phi$ magnitude
does not decay with distance from the interface.  M$^+$ displacement
or the transfer of an $e^-$ following an electrochemical reaction changes
$d$, $\Phi$, and hence ${\cal V}_e$ via Eq.~\ref{eq3} in finite simulation
cells.  Under constant voltage conditions, $\Delta {\cal V}_e$
should be zero over the course of a reaction.  To achieve this
condition, constant ${\cal V}_e$ ensembles\cite{peterson} or extrapolation
to large cell sizes\cite{iceland} have been applied in electrocatalysis
calculations.  Related methods have recently been applied to battery
interfaces.\cite{otani_batt}

In some cases, apparent degradation or ``disordering'' reactions can extend
to micron depth beyond the interface\cite{lipon_lco} --- far beyond DFT
length scales.  Fitting reactive force fields can extend the time- and
length-scale associated with the battery interface reaction
zone.\cite{bedrov_reaxff} However, spin-dependent reactive potentials have not
been devised for transition metal oxides.  Standard modelling
protocols like the ``master equation'' exist to deal with kinetics-controlled,
multistep reactions in catalysis and gas phase reactions.\cite{kinetics}  So
far the complexity of battery reactions has precluded their use.  

To avoid challenging prospect of calculating all possible reaction mechanisms,
whenever possible, we have focused on elucidating the stability of proposed
surface film components.\cite{batt}  We call this the ``surface film
instability paradigm.''  Recent application of artificial intelligence to
determine locally optimal interfacial atomic structures\cite{tateyama2} can
yield interfaces stable against bond-breaking.  This promising approach may 
efficiently satisfy our stability criterion if quasi-kinetic constraints 
can be added.

\color{black}
\subsection{Non-DFT and non-planewave Methods}

We briefly mention non-DFT electronic structure methods which are not our
focus.  Quantum mechanics/molecular mechanics (QM/MM) methods have accelerated
solution phase reactions by confining DFT to a small ``QM'' spatial
region.\cite{qmmm1,qmmm2}  Since electrochemical calculations involve voltages
which require metallic electrodes, the ``QM'' required may still be
substantial.  Tight-binding methods have been long proposed for aqueous phase
electrochemistry but require more development; tight-binding treatment
of organic species and materials associated with battery o-SEI has yet
to be developed.  We apply a plane-wave plus projector-augmented wave (PAW)
basis set, but the discussions above are applicable to all DFT basis
sets, including mixed-planewave/atomic basis.\cite{siepmann}
\color{black}

\subsection{DFT Details}

All DFT calculations are conducted under T=0~K ultrahigh vacuum (UHV)
condition, using periodically replicated simulation cells and the Vienna
Atomic Simulation Package (VASP) version 5.3.\cite{vasp1,vasp1a,vasp2,vasp3}
A 400~eV planewave energy cutoff and a 10$^{-4}$~eV convergence criterion
are enforced.  Most calculations apply the PBE functional.\cite{pbe}  In
one case, HSE06 is used as a spot check.\cite{hse06a,hse06b,hse06c}   The
standard dipole correction is applied.\cite{dipole}

We consider Li metal (001)/LiF (001) interfaces as exemplars of
Fig.~\ref{fig2}b.  The base model has a 17.28$\times$17.28$\times$36~\AA$^3$
simulation cell with stoichiometric Li$_{294}$F$_{144}$.  We conduct a 10~ps
AIMD trajectory at T=400~K to equilibrate the interface, and then optimize the
final MD configuration.  The bottom two Li metal layers are held fixed
throughout.  When an O$_2$ molecule is added to the surface, the $c$ lattice
constant is expanded by 4~\AA.  2$\times$2 Brillouin sampling is used.
For the purpose of checking PBE results with the HSE06 functional in this
model, $\Gamma$-point sampling is employed; decreasing the $k$-point grid
changes ${\cal V}_e$ by less than 0.12~V.

To make connection with corrosion research, Al (111)/$\alpha$-Al$_2$O$_3$
(0001) interfaces are also considered.  No definitive metal-oxide
interfacial structure is known in the experimental literature,
and proposed DFT-configurations with subtle differences yield significant
changes in $\Phi$.\cite{bredas,wang}  This is unlike the weakly interacting
and more robust Li-metal/LiF interface.  Like Ref.~\onlinecite{costa},
we apply AIMD simulations to equilibrate this interface.  Our AIMD is conducted
on a primitive surface cell with dimensions 4.81$\times$8.33$\times$50~\AA$^3$
and Al$_{70}$O$_{54}$(H$_2$O)$_6$ stoichiometry.  
\color{black}
This is the size of a single surface unit cell, chosen to be commensurate with
those of Refs.~\onlinecite{bredas} and~\onlinecite{wang}, so that a direct
comparison of the respective properties with those work can be made.  We aim
to provide more relaxation than the purely T=0~K configuration relaxation of
these two work.  Our cell size is smaller than that in Ref.~\onlinecite{costa}.
We do not claim our procedure yields  the definitive structure of this
complex interface, which likely depends on material preparation conditions.
\color{black}

Unlike previous DFT work,
we add 3~H$_2$O atop each surface Al$^{3+}$ in an attempt to make surface
Al ions 6-coorindated.  After a 0.8~ps AIMD trajectory at T=400~K, the
system is cooled to T=100~K in 0.15~ps.  We find that one-third of the added
water evaporates/detaches from the surface; they are removed.  Another third
dissociates into H$^+$ and OH$^-$.  The atomic configuration is then optimized
at T=0~K, resulting in a configuration where each surface Al is 4-coordinated,
and the intact H$_2$O per surface Al lies parallel to the surface, coordinated
only to other surface O or OH groups.  A 3$\times$2 surface supercell is
constructed from this primitive cell and is applied to examine oxygen vacancies
with 2$\times$2$\times$1 Brillouin sampling.  

The LiOH (001)/Li (111) interface is examined for kinetics purposes.
Here the base model has a 13.53$\times$14.33$\times$40~\AA$^3$ simulation
cell and a Li$_{224}$O$_{06}$H$_{96}$ stoichiometry, and 
2$\times$2$\times$1 Brillouin sampling is applied.  In all cases Li and Al
metals, softer than fluorides and oxides, are strained to match the fluoride or
oxide supercell lattice constants.  This changes the pure metal work function
by $\sim$0.1~V.  All simulation cells are charge neutral.

\section{Results}

\subsection{Lithium Metal in Vacuum}

Consider a BCC lithium solid cleaved in the (001) direction in vacuum.
The cohesive energy is that of Li-metal, and ${\cal V}_i$=0.00~V
vs.~Li$^+$/Li as we remove one entire layer of Li atoms, according to
Eq.~\ref{eq1}.  (The small strain energy is neglected.) The PBE-predicted
work function is $\Phi$=3.02~eV, close to the experimental value of
2.93~eV.\cite{crc}  From Eq.~\ref{eq2}, ${\cal V}_e$=1.65~V vs.~Li$^+$/Li(s).
Hence the system is out of $i$-equilibrium, at overpotential against Li
dissolution.  If there were a Li reservoir and an electrolyte, and
${\cal V}_e$=1.65~V were enforced by a potentiostat, the Li slab would
completely dissolve.

\begin{figure}
\centerline{\hbox{ \epsfxsize=4.00in \epsfbox{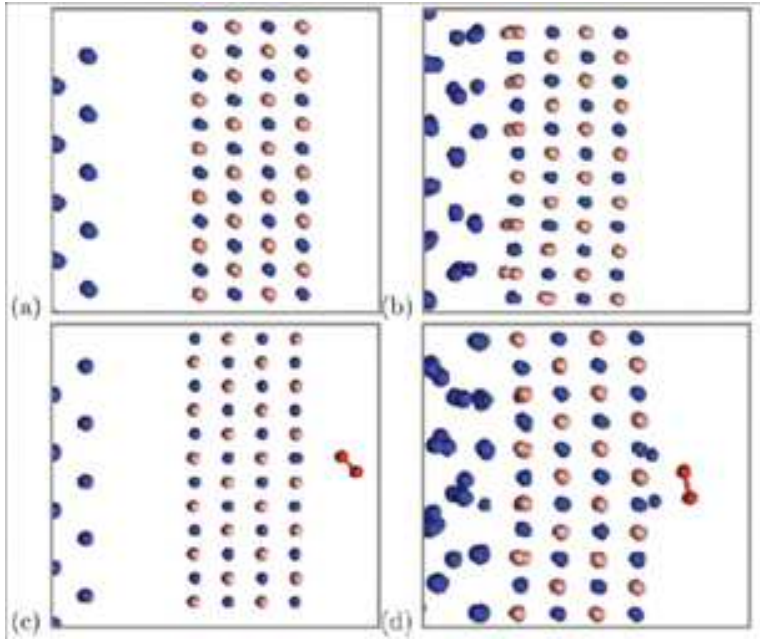} }}
\caption[]
{\label{fig3} \noindent
(a) Li (001) and LiF (001) slabs separated by $\sim$5~\AA;
(b) Li (001) and LiF (001) slabs in contact;
(c)-(d) optimized configurations of panel (a) when O$_2$ is added, and
of panel (b) when 9 interlayer Li atoms and a O$_2$ are added.
Blue, pink, and red spheres depict Li, F, and O atoms, respectively.
The $+z$ direction is to the right.
}
\end{figure}

\begin{figure}
\centerline{\hbox{ (a) \epsfxsize=4.00in \epsfbox{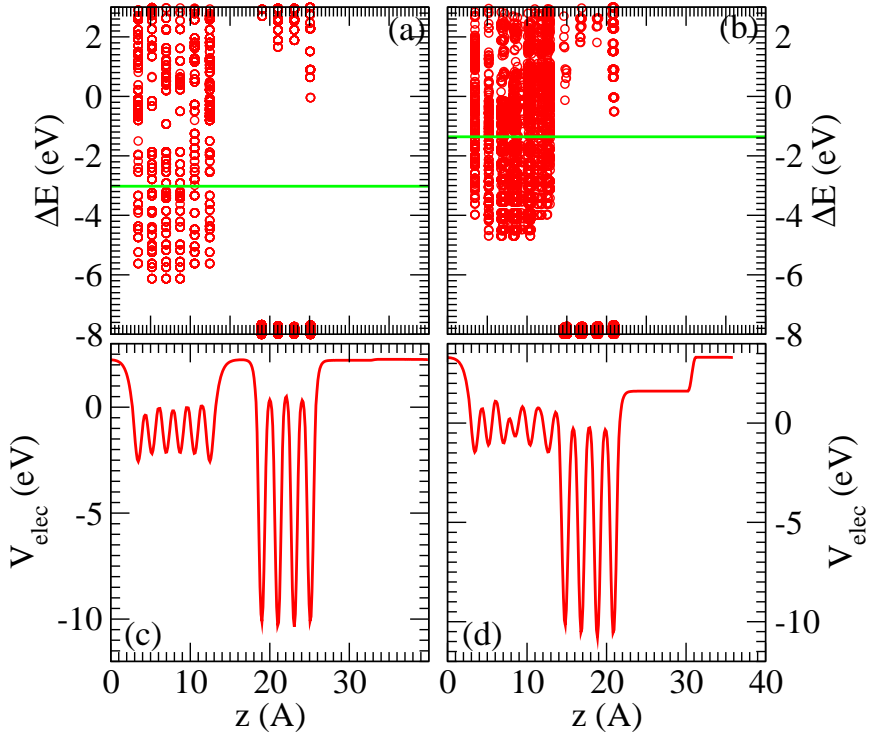} }}
\caption[]
{\label{fig4} \noindent
(a)-(b) Local densities of state of valence electrons
associated with Figs.~\ref{fig3}a-b. Red
circles depict Kohn-Sham orbitals on atoms which contribute to the wavefunction
by more than 2\%; the green line are Fermi levels and vacuum is at
$\Delta E$=0.0~eV.  (c)-(d) Electrostatic potentials associated with
Figs.~\ref{fig3}a-b, respectively; the left and right side vacuum levels
correspond to coated and uncoated Li metal, separated by a discontinuity
due to the artificial dipole layer.\cite{dipole}
}
\end{figure}

Fig.~\ref{fig3}a reinforces this conclusion.  A 8~\AA\,-thick LiF slab with its
(001) facet exposed is located 5~\AA\, apart from the Li (001) slab.  Neither
the Li nor the LiF slab has a net charge or a net dipole moment when isolated.
Fig.~\ref{fig4}a depicts $z$-axis-resolved orbital positions showing no
electronic overlap between the slabs separated by vacuum.  The electrostatic
potential profile (Fig.~\ref{fig4}c) suggests that the 5~\AA\,-separated slabs
remain largely dipole free.  ${\cal V}_e$=1.65~V, only weakly perturbed from
that of isolated Li-metal slab.  Once again, ${\cal V}_i$$\neq$${\cal V}_e$.

This finding cannot be over-emphasized: {\it it is incorrect to assume
that Li metal is always at $i$-equilibrium}, i.e., at 0.0~V vs.~Li$^+$/Li(s),
in an interfacial simulation cell. Experimentally, lithium is stable at
{\it or below} 0.0~V; when stripping Li under overpotential conditions, Li
also exists above 0.0~V.  No conclusion about ${\cal V}_e$ can be drawn from
the existence of Li metal in a simulation cell.  The same is true of other
electrodes, including intercalation and conversion electrodes.
The overpotential associated with plating and
stripping of Mg anodes can approach~1~V.\cite{aurbach_mg}  Au electrodes are
featured in numerous DFT interfacial simulation cells.\cite{arias}  Reported
voltages (${\cal V}_e$) in that literature are correctly pegged to 
($-E_{\rm F}$) modulated by the electric double layer,\cite{arias} not assumed
to be at the Au$^{3+}$+3$e^-$$\rightarrow$Au(s) half cell voltage.
Distinguishing ${\cal V}_e$ and ${\cal V}_i$ establishes electrochemical
equilibrium and permits the modelling of overpotential effects.

\subsection{Surface Films Strongly Affect ${\cal V}_e$}

Fig.~\ref{fig3}b depicts the configuration obtained after we put the 
LiF and Li surface slabs in contact.  The system develops a large dipole
moment, and $\Phi$ drops from the vacuum value of 3.02~eV to 1.54~eV.
${\cal V}_e$ becomes 0.17~V (Eq.~\ref{eq2}), much closer to ${\cal V}_i$=0.0~V
than bare Li (001).  The surface film brings it close to $i$-equilibrium ---
without adding surface charges or liquid electrolyte.  Similar, significant
``contact potential'' drops upon adding an inorganic film, which reduces
$\Phi$ (and hence ${\cal V}_e$ via Eq.~\ref{eq2}), have been noted in the
corrosion,\cite{marks} electronics,\cite{kondo},
solid electrolyte,\cite{solid,maier} and Li-air
battery\cite{norskov} modelling literature.  They are partly caused by $e^-$
density rearranging within each material at the interface, creating a dipole
layer (Eq.~\ref{eq3}).  The reduction in $\Phi$ is reminiscent of the effect
of a water monolayer on inert metal surfaces,\cite{gross} except that the LiF
film exhibits an induced, not permanent, dipole moment.  Significant contact
potentials can also be inferred from experiments,\cite{voltage_map,kpfm}
although they are conducted at less than atomic resolution.  

We can further reduce ${\cal V}_e$ to -0.01~V in Fig.~\ref{fig3}b
($i$-equilibrium condition) by manually adding an interlayer of Li atoms
directly underneath some of the F$^-$ anions at the interface.\cite{gb}
Adding Li atoms in this way is found to be thermoneutral to within 0.05~eV 
after accounting for Li chemical potential (Eq.~\ref{eq1}).  It suggests
that the energy/voltage landscape at the interface has many similar
local minima.  This interlayer approach, which reflects the grand
canonical ensemble environment, is motivated by the ``interfacial charge
storage'' idea.\cite{maier}  Fig.~\ref{fig4}b and~\ref{fig4}d depict the 
resulting orbital alignment and electrostatic potential.  A finite band gap
remains in the LiF region, and no LiF surface state is occupied.  

Doubling the LiF slab thickness to 18~\AA\, without the Li-interlayer atoms
yields a $\Delta \Phi$ slightly larger by 0.05 eV, leading to
${\cal V}_e$=0.22~V.  The somewhat different ${\cal V}_e$ partly reflects
system size effect, but is likely within the uncertainty margin.  The
Li$_2$O (111)/Li (001) interface yields a contact
potential drop of roughly the same size; however, there is small amount of
$e^-$ leakage to the outer Li$_2$O surface,\cite{gb} which in reality would
have led to further passivation by the organic materials outside. 
The PBE functional used herein has
delocalization errors\cite{wtyang} which may permit unphysical $e^-$ leakage
from Li metal into LiF.  To check that this does not overestimate the contact
potential, we also apply the more accurate HSE06 functional to the
Li/8-\AA-LiF interface.  The resulting $\Delta \Phi$ is in fact 10\% larger
than the PBE prediction, showing that our PBE calculations most likely do
not exaggerate the contact potential drop.

The flatness of the valence band edge (Fig.~\ref{fig4}b) emphasizes that we
are at ``potential-of-zero-charge''-like conditions.  Our focus is the
near-anode region, not the voltage profile over long length scales
(Fig.~\ref{fig1}b).  Nevertheless, if in real-life batteries the lithium
anode behaves like Fig.~\ref{fig4}b, the voltage profile in Fig.~\ref{fig1}b
would remain flat in the liquid electrolyte region until it reaches the
cathode EDL.  This ignores the fact that LiF is not the only SEI component.
Charges at defect sites and other factors may also introduce
electric fields near the anode surface and creates an EDL, part of which
may now reside at the solid-solid interface.  The exact partitioning of solid
state vs.~liquid state EDL depends on the free energy cost of developing those
EDL's.  This little-explored EDL partitioning offers a potential design
principle for improving battery interfaces.  

\subsection{How $i$-Equilibrium and ${\cal V}_e$ Affect DFT Predictions}

${\cal V}_e$ is seldom reported or controlled in DFT modelling of battery
interfaces.  Some reported DFT interface simulations are likely out of
$i$-equilibrium;\cite{budzien,ucsd} they are at as-synthesized rather than
battery cycling conditions with well-defined voltages. What are the
consequences?

${\cal V}_e$ is crucial for determining the electrochemical equilibrium
conditions where M$^+$ insertion from liquid electrolyte into metallic
electrode is not at an overpotential.\cite{voltage,otani_batt,tb}
The properties under these conditions, e.g., M$^+$ insertion kinetics and
barrier, is expected to strongly depend on ${\cal V}_e$.  ${\cal V}_e$
is also crucial for passive electrodes acting as $e^-$ emitter.  The
correct ${\cal V}_e$ value yields redox event onset at the right voltage.  In
this light, the edge-termination dependence of LiC$_6$ on electrolyte redox
behavior in our early work\cite{budzien} is likely the result of the graphite
edge sites giving different ${\cal V}_e$, not computed at that time.

Fig.~\ref{fig3}c-d qualitatively illustrate this effect by placing a O$_2$
molecule on to the outer LiF surfaces of Fig.~\ref{fig3}a and~\ref{fig3}b.
This example draws on DFT modelling of O$_2$ reduction on Al$_2$O$_3$-coated Al
surfaces,\cite{costa} relevant to the growth of a passivating film there, and
sidesteps more complex organic molecule redox demonstrations.\cite{vedge,gb}
When LiF is coated on Li with an ``interlayer Li'' (Fig.~\ref{fig3}b),
${\cal V}_e$=0.0~V, and an $e^-$ readily transfers through the LiF film to
the adsorbed O$_2$.  Bader analysis\cite{bader} confirms that a -1.00$|e|$ net
charge exists on the oxygen species, i.e., it is a superoxide (O$_2^-$).  When
the Li and LiF slabs are well-separated (Fig.~\ref{fig3}a), ${\cal V}_e$ is
much less reductive (1.67~V), and the O$_2$ weakly physisorbs instead.
Magnetic moment analysis suggests that O$_2$ is a triplet molecule.  Bader
analysis yields a -0.31~$|e|$ net charge.  The non-integer value is
unphysical and is likely due to DFT/PBE localization error\cite{wtyang} for
this space-separated configuration, which is challenging for DFT methods.

In the literature, it has been reported LiF-coated Li metal rapidly decomposes
EC molecules in AIMD simulations.\cite{borodin_lif,bal_lioh}  Fig.~\ref{fig3}
suggests the reason is that LiF lowers ${\cal V}_e$ to values which initiate
electrochemical reduction.  On bare Li metal surfaces, ${\cal V}_e$ is
significantly higher.  Despite this, many organic species also rapidly
decomposes on bare Li in AIMD simulations.\cite{goddard,nasa} But these
reactions are likely chemical, featuring reactants in contact, rather than
electrochemical (i.e., long range $e^-$ transfer) in nature.   We will return
to this theme in Sec.~\ref{field_effect}.

${\cal V}_e$ and electric fields associated with it should affect charged
carriers transport properties.\cite{santosh} Here we use a negative example
and examine how Li$^+$ vacancy motion towards the metallic electrode affect
computed voltages.  Again we consider a Li metal slab as in Fig.~\ref{fig3}b,
with a LiF layer with twice the thickness ($\sim$18~\AA\, thick, not shown).
Without Li-vacancy, ${\cal V}_e$=0.19~V.  With a Li-vacancy 10~\AA\, from
the lithium metal surface, $d$ increases and ${\cal V}_e$ rises to 0.98~V.  
When the vacancy is 4~\AA\, away from the surface, ${\cal V}_e$=0.54~V,
still higher than the no-vacancy configuration.  Hence the system is not
at constant potential during Li-vacancy transport.  This is an artifact of
the finite surface area of the simulation cell (Eq.~\ref{eq3}).
The need to move atoms while keeping ${\cal V}_e$ constant is known in
electrocatalysis modelling.\cite{rossmeisl13,otani,peterson,iceland} Battery
calculations generally have larger lateral surface areas than electrocatalysis
and smaller $\sigma$ (Eq.~\ref{eq3}), but $d$ can be much larger unless M$^+$
moves in a high-dielectric liquid electrolyte\cite{voltage,vedge,otani_batt}
instead of a relatively low dielectric solid film.  In general, if there 
are large changes in ${\cal V}_e$ along a Li diffusion pathway, substantial
corrections may be needed to recover the constant potential energy landscape.

\subsection{Relation to Atmospheric Corrosion}

\begin{figure}
\centerline{\hbox{ \epsfxsize=4.00in \epsfbox{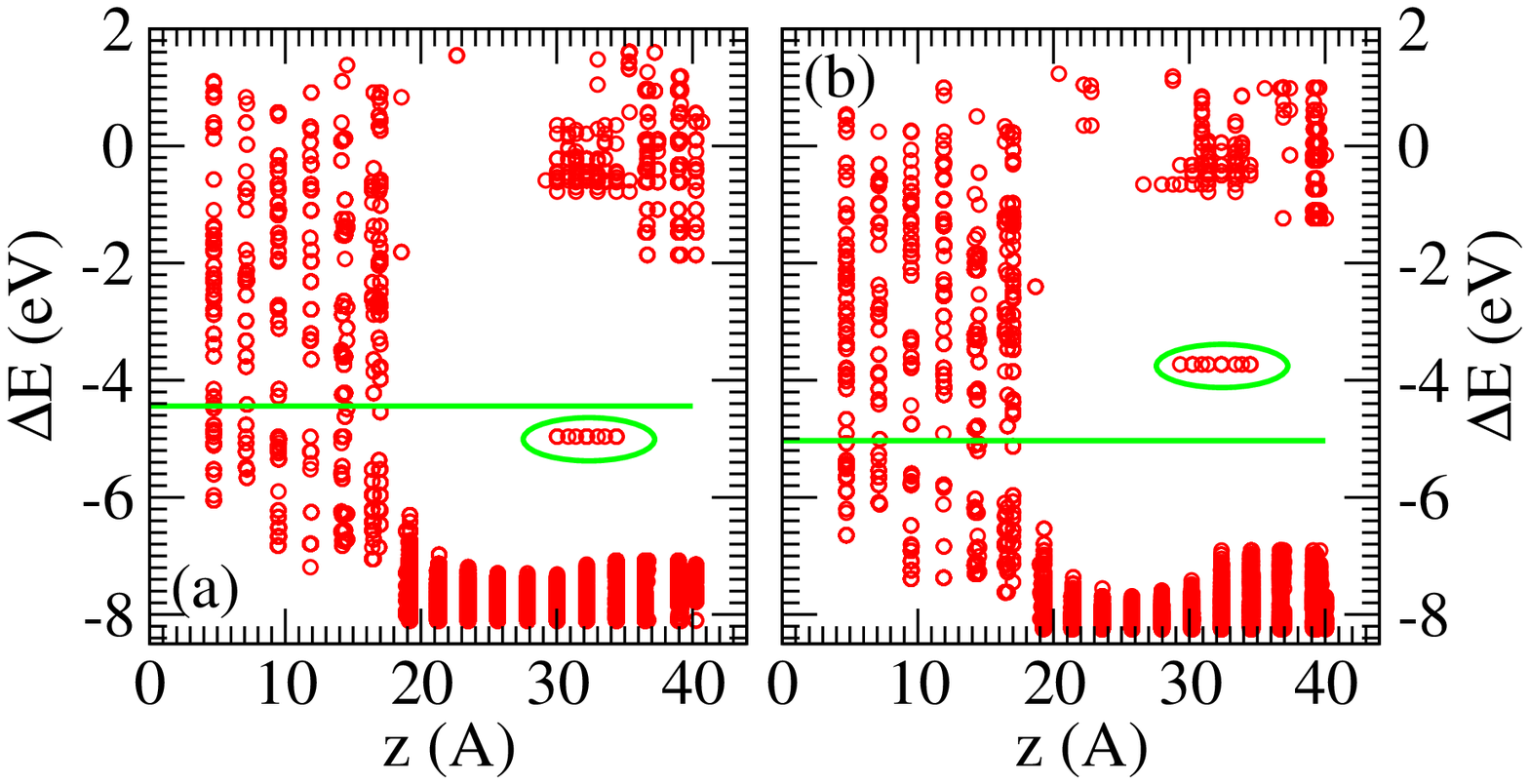} }}
\centerline{\hbox{ \epsfxsize=4.00in \epsfbox{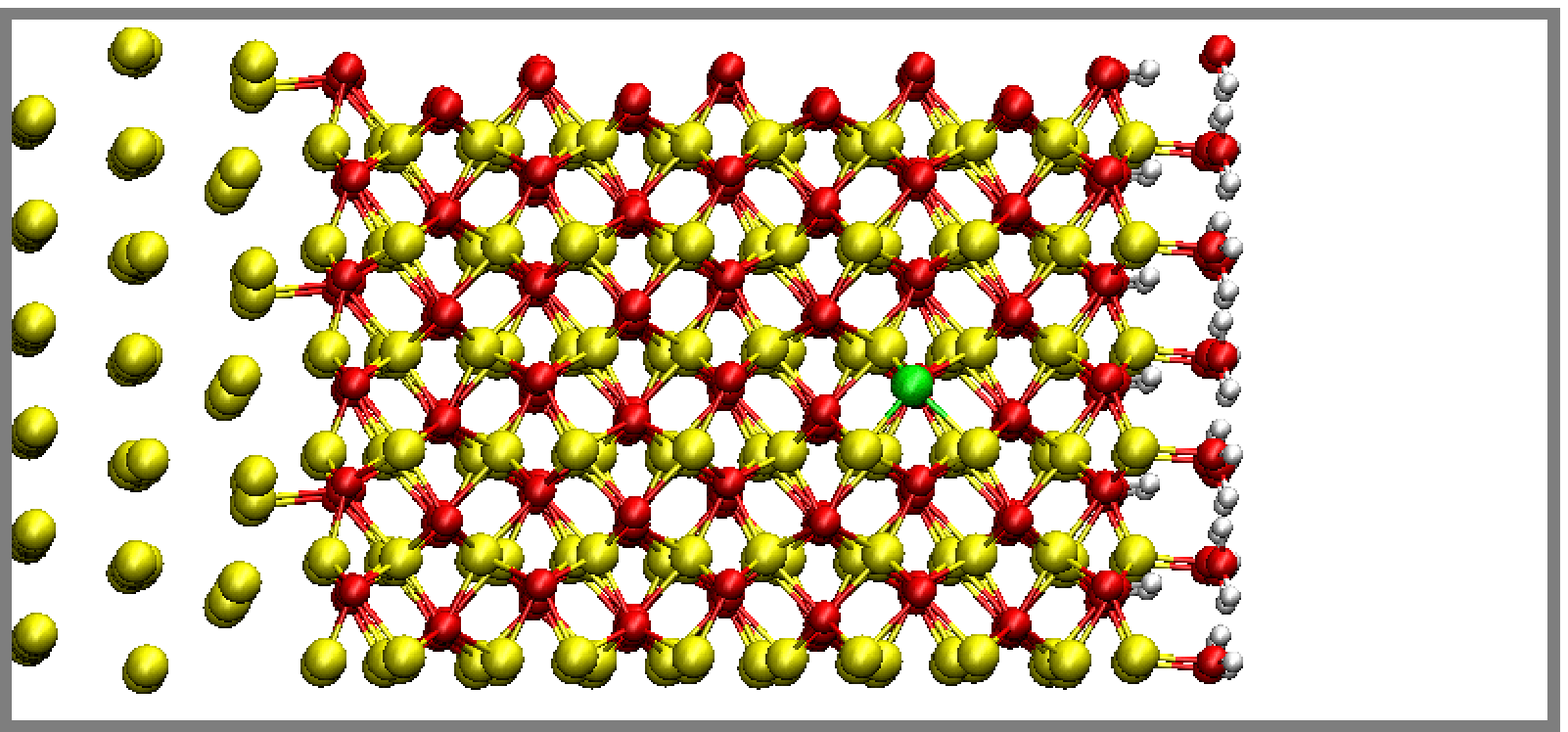} }}
\caption[]
{\label{fig5} \noindent
(a) Local densities of state of valence electrons
associated with the Al(111)/$\alpha$-Al$_2$O$_3$
(0001) interface.  The occupied states in the gap which correspond to a
neutral oxygen vacancy are circled in green.  (b) Same as (a), but with 2
H$^+$ vacancies at the vacuum interface to create an electric field.
(c) Atomic configuration of the slab.  The green O-atom at $z$=32.2~\AA\, is
removed to create a vacancy.  Yellow spheres are Al atoms.
}
\end{figure}

Next we consider contact potential and electric field effects on the occupation
of orbitals associated with defects.  In battery modelling, related studies
have been treated mostly at a flat-band level without explicitly considering
electrified interfaces.\cite{tian2019,gb} Because more prior relevant work
exists in the Al$_2$O$_3$ literature, we use the Al(111)/$\alpha$-Al$_2$O$_3$
(0001) interface\cite{costa} relevant to atmospheric pitting corrosion of
aluminum, as example.  Of particular interest is the charge state of oxygen
vacancies.  The ``point-defect model'' (PDM),\cite{macdonald} a widely used
approach for predicting passivating oxide breakdown that leads to pitting,
proposes doubly positively charged O-vacancies.  When Al metal is present in
a DFT simulation cell, the charges of defect states are determined by
$E_{\rm F}$ and ${\cal V}_e$.

Fig.~\ref{fig5}a depicts the orbital alignment of this interface with
an oxygen vacancy at position shown in Fig.~\ref{fig5}c.  By comparing with
a slab without vacancies, we identify orbitals circled in green as those
associated with the O-vacancy.  They lie below the Fermi level $E_{\rm F}$,
indicating that the vacancy is uncharged.  Consistent with this, the valence
band edge is relatively flat, showing little electric field.
${\cal V}_e$=0.16~V.

To recover positively charged oxygen vacancies, we remove H atoms from two of
the twelve surface AlOH groups.  As expected, DFT predicts that the H-vacancies
carry negative charges, which should be compensated by positive charges on Al
metal surfaces in the charge-neutral simulation cell.  This raises ${\cal V}_e$
to a more oxidizing 0.59~V, and raises the local valence band edge near the
surface (Fig.~\ref{fig5}b).  Now creating the same O-vacancy as before yields
unoccupied orbital levels above $E_{\rm F}$ (Fig.~\ref{fig5}b), and a doubly
positively charged defect which dovetails with PDM\cite{macdonald} emerges.
A more significant electric field can be inferred from the valence band edge
in the range 24$<$$z$$<$30~\AA.

The O-vacancy is periodically replicated in Fig.~\ref{fig5}c.  In reality,
a spatial distribution of O-vacancies should exist over a range of $z$ values.
In pitting corrosion, adsorbed anions are likely to play the electrostatic
role of negatively charged H vacancy in Fig.~\ref{fig5}c.  Adsorbed halide ions
are indeed strongly linked with pitting corrosion onset.  The ${\cal V}_e$'s
constructed for these slabs are too high compared with Al pitting onset
voltages.  Nevertheless, this slab model appears a promising platform
for investigating voltage effects in Al$_2$O$_3$ passivating films.
Another fruitful connection between battery and corrosion is the possible
galvanic lithium corrosion.\cite{cui2019}

\subsection{The Li(s)/LiOH Interface, a Kinetics Case Study}
\label{kinetics}

To illustrate the kinetics-controlled nature of battery interfaces, we
consider lithium hydroxide on lithium metal surface.  LiOH has occasionally
been cited as a SEI component, both experimentally and
theoretically.\cite{aurbach_two_layer,bal_lioh} In Ref.~\onlinecite{bal_lioh},
AIMD simulations show that organic solvent decomposes on LiOH (010)-coated Li
metal (001) surfaces; the LiOH film itself does not react.  However, OH groups
are expected to give off H$_2$ gas at low voltages.\cite{aurbach_two_layer}
Here we explicitly re-examine LiOH stability.

First we consider thermodynamics.  At T=0~K, both
\begin{eqnarray}
{\rm 2 LiOH(s)} + {\rm 2 Li(s)} &\rightarrow& {\rm 2 Li}_2{\rm O(s)} 
	+ {\rm H}_2{\rm (g)} \label{eq6}, {\rm and} \\
{\rm LiOH(s)} + {\rm 2 Li(s)} &\rightarrow& {\rm Li}_2{\rm O(s)}
	+ {\rm LiH(s)} \label{eq7}
\end{eqnarray}
are exothermic, by -1.74 and -1.70~eV, respectively.  The reaction
\begin{equation}
{\rm H}_2 {\rm (g)} + {\rm 2 Li(s)} \rightarrow {\rm 2 LiH(s)} \label{eq8}
\end{equation}
is also exothermic by -1.65~eV.  H$_2$ gas release is favored by a $\sim$0.4~eV
entropy gain at room temperature over that at T=0~K, but this is insufficient
to reverse the exothermicity of Eq.~\ref{eq8}.  LiH\cite{aurbach_two_layer} has
indeed been observed using cryo-TEM,\cite{cryo-tem} although its presence
may depend on electrolyte water content.\cite{cryo-tem1}
Thermodynamically, LiOH solid is predicted to be unstable against lithium
metal.  Kinetically, both the LiOH film and isolated OH$^-$ units, which come
from reaction with water, are stable within $<$100~ps AIMD trajectory
lengths.\cite{bal_lioh}  Here we turn to activation barrier ($\Delta E^*$)
calculations to extrapolate to longer time scales.

\begin{figure}
\centerline{\hbox{ \epsfxsize=4.00in \epsfbox{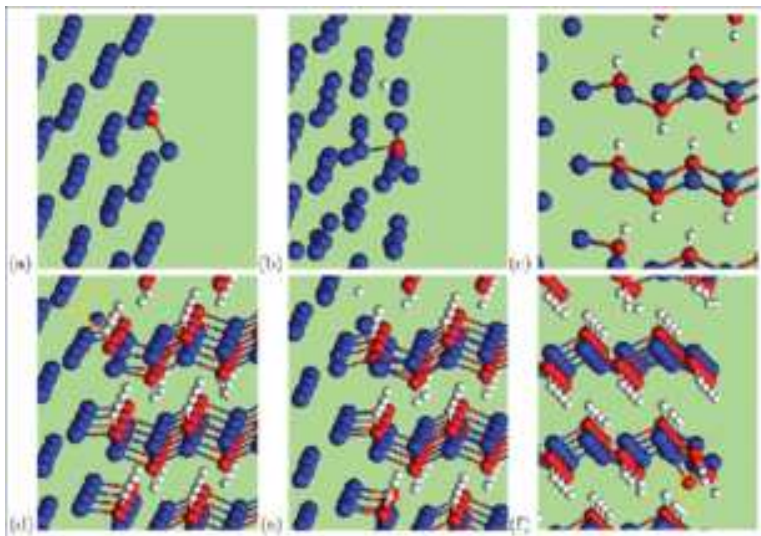} }}
\caption[]
{\label{fig6} \noindent
(a)-(b) Intact and decomposed LiOH unit on Li (001).
(c) LiOH (010) slab on Li (001) surface focusing on interface region, 
showing all OH groups there coordinated to surface Li atoms.
(d) Adding an extra Li (circled in yellow) at the interface.  The arrow
   indicates the H atom to be transferred.
(e) Same as (d), but with a hydrogen transferred to the surface.
(f) Top of LiOH (010) slab with removal of a H atom from within yellow circle.
The $+z$ direction is to the right.
}
\end{figure}

Fig.~\ref{fig6}a depicts an isolated LiOH unit optimized on the Li (001)
surface.  In Fig.~\ref{fig6}b, the proton is detached to the Li surface,
with a favorable $\Delta E$=-1.29~eV released.  NEB calculations reveal that
$\Delta E^*$=0.72~eV.  From Eq.~\ref{eq0}, this translates into an average
reaction time of 0.82~s. This is fast on battery time scales, but too slow
to be observed in unconstrained AIMD simulations ($<$10$^{-9}$s in duration).
We conclude that isolated LiOH is kinetically unstable on Li (001).

Next we consider the Li (001)/LiOH (010) interface.  Its proposed existence
assumes that adsorbed LiOH monomers can nucleate in sub-second time scales
before they decompose.   The lattice matching is such that all OH groups
at the interface are coordinated to surface Li metal atoms (Fig.~\ref{fig6}c),
unlike LiF and Li$_2$O.\cite{gb} Breaking one of the three inequivalent OH
bonds to form a H$^-$ on the lithium metal surface releases
$\Delta E$=-0.79~eV, favorable for reaction.  The barrier is
$\Delta E^*$=1.27~eV, which indicates a reaction exceeding experimental time
scales.  The other two inequivalent OH groups at the interface exhibit
$\Delta E$=-0.60~eV but $\Delta E^*$ are similar at 1.20~eV. If one were
to deposit Li vapor on defect-free LiOH (010), reactions are expected only
at elevated temperatures.

\begin{figure}
\centerline{\hbox{  \epsfxsize=4.00in \epsfbox{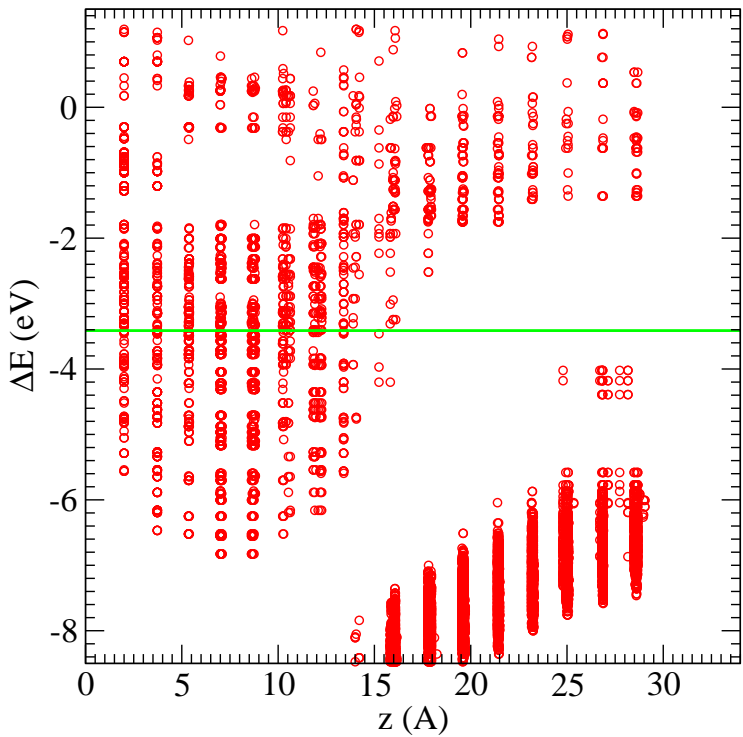} }}
\caption[]
{\label{fig7} \noindent
Local densities of state of valence electrons
associated with Fig.~\ref{fig6}f with surface H
vacancy.  For color key, see Fig.~\ref{fig4}.
}
\end{figure}

However, the above calculations ignore the grand canonical nature of the
interface during cycling.  A metastable configuration can be created by adding
a Li atom to the surface (Fig.~\ref{fig6}d).  $\Delta E'$ is 0.22~eV after
accounting for lithium chemical potential.  From this configuration,
breaking the O-H bond (Fig.~\ref{fig6}e) releases $\Delta E$=-1.07~eV
and $\Delta E^*$=0.76~eV.  Accounting for $\Delta E'$ but not zero point
energy (ZPE) that further reduces $\Delta E^*$ in proton transfer
reactions,\cite{tuckerman} the energy at the transition state is 0.98~eV above
that of Fig.~\ref{fig6}c.  From Eq.~\ref{eq3}, this is one reaction per
8~hours at T=300~K.  Without mapping out all possible reaction paths
via the master equation,\cite{kinetics} DFT/PBE calculations predict at least
one pathway whereby that a pristine LiOH (010) slab should react with lithium
metal on an experimentally observable time scale, likely forming Li$_2$O
on Li metal surfaces.  We conclude that the LiOH SEI component seen in
experiments are likely not in contact with Li metal.  It is more likely lodged
on other, more stable SEI components such as Li$_2$O.

Thus AIMD simulations of liquid or solid electrolyte decomposition
on Li metal surfaces are extremely useful for giving qualitative insights
about possible reaction mechanisms, especially at the initial stages of
interfacial degradation reactions.  But they currently cannot rule out
that reactions can occur at $>$1~ns time scales.

\subsection{Voltage and Electric Field Effects on Kinetics}
\label{field_effect}

This LiOH (010) example permits an in-depth exploration of the effect of
${\cal V}_e$ on $\Delta E$ and $\Delta E^*$ associated with parasitic
reactions.  Fig.~\ref{fig6}c, d,~and~e are consistent with ${\cal V}_e$=0.42~V,
0.46~V, and 0.45~V, respectively.
Unlike the LiF interface (Fig.~\ref{fig3}a), all O~atoms at the surface are
coordinated to Li metal atoms, and attempts at adding an interlayer like in
Ref.~\onlinecite{gb} fail to lower ${\cal V}_e$.  To achieve $i$-equilibrium
(${\cal V}_e$=0.0~V) requires cations outside the LiOH slab to reduce $d$ via
Eq.~\ref{eq3}.  Here we take an opposite approach.  Motivated by our
Al$_2$O$_3$ analysis (Fig.~\ref{fig5}), we create a H vacancy at the LiOH
outer surface (Fig.~\ref{fig6}f). As expected, DFT calculations yield a
negatively charged H-vacancy in the neutral cell, which makes $d$ more
positive in Eq.~\ref{eq3} and increases ${\cal V}_e$ from 0.42~V to 2.17~V while
${\cal V}_i$ remains 0.0~V.  The system is now at a significant overpotential
with respect to Li dissolution from the anode.  The valence band edge shown
in Fig.~\ref{fig7} suggests a significant electric field of $\sim$0.22~V/\AA.

Remarkably, $\Delta E$ and $\Delta E^*$ are almost unchanged. They are now
-1.09~eV and +0.77~eV, within 0.02~eV of the values at ${\cal V}_e$=0.46~V.
This is despite the fact that Bader charge analysis suggests the H transferred
to the surface now has a $-2.1~|e|$ charge (i.e., it is a hydride).  In
electrocatalysis modelling, $|e$| times changes in ${\cal V}_e$ (designated
as ``$U$'') is often added post-processing to $\Delta E$ on metal surfaces in
reaction steps that involve $e^-$ transfer, and $\Delta E^*$ have larger
field-dependence.\cite{chan}  In Fig.~\ref{fig6}, an added $\pm|e|$${\cal V}_e$
should emerge in $\Delta E$  only if a charged species transverse the entire
solid state EDL.  The lack of change in $\Delta E^*$ is however unusual, and
may reflect strong metal surface screening.

Other parasitic reactions on Li metal we have considered, like those at the
Li/LiPON\cite{lipon} (oxy-phosphorus nitride inorganic polymers) and
Li/Li$_2$CO$_3$\cite{batt} interfaces, exhibit similarly small voltage
dependences until the field strengths imposed by localized defects are
sufficient to initiate different degradation mechanisms that involve long-range
$e^-$ transfer.\cite{batt} Experimentally, it is known that storage
(without applied voltage) and cycling conditions (applying voltages) yield
different behavior in metal anodes; the dendrite growth in particular depends
on charging rate and overpotentials applied.\cite{gross1}  If our anecdotal
DFT evidence is true in general, it suggests that the difference between
storage and cycling conditions does not stem from primary reaction rates, but
originates from long range $e^-$ transfer to defects which may occur in unison
with strictly chemical reactions at the interface, and/or Li$^+$ transport
effects.  At other electrode interfaces, electric field- and voltage- effects 
remain to be distinguished and elucidated.  

\section{Conclusions and Outlook}

In this work, we apply simple examples with explicit model electrode/surface
film interfaces to illustrate some key challenges
specific to electronic structure DFT modelling of interfaces in high energy
density batteries.  These include voltage calibration/control, the thin-film
covered (``dirty'') nature of electrodes in monovalent cation (M$^+$) batteries
where M=Li or Na, 
and the fact that interfaces are kinetics-controlled.  We find that contact
potentials at metal-solid electrolyte interfaces, which can only be computed
with explicit interfaces in DFT models, cannot be neglected in
understanding the voltage profile across interfaces.  The ionic (``open
circuit'') voltage ${\cal V}_i$, which implies equilibrium conditions, and
electronic (``instantaneous'') voltage ${\cal V}_e$, must be distinguished.
These two quantities determine whether interfaces are at equilibrium, and
enable modelling of interfaces at overpotential conditions.  ${\cal V}_e$
is shown to govern long range transfer and electric fields that modify
transport properties.  It appears to play little role in chemical reactions
where reactants are in direct contact with lithium metals anodes.  However,
the roles of voltage and electric fields on reactions at other battery
interfaces need further clarification.

The surface-film covered nature of electrodes is not an inconvenience; it
is a major part of the interfacial physics of batteries, and should be
regarded as a new basic science paradigm and opportunity.  In contrast,
modelling pristine metal electrode surfaces of M$^+$ batteries, 
is mainly relevant to the initial stages of battery assembly.
Unconstrained {\it ab initio} molecular dynamics (AIMD) simulations cannot
conclusively prove the stability of interfacial components because trajectory
lengths are 12~orders of magnitude shorter than battery operational time
scales; barrier finding computational techniques are needed to understand
kinetics-controlled interfaces.  The M$^+$ content across the entire interface
follows the grand canonical ensemble.  Cathode interfaces are challenging
because of the polaron-conducting nature of cathode materials.  Borrowing from
other branches of computional electrochemistry like corrosion, supercapacitors,
and electrocatalysis presents good opportunities for advancing the
current state of battery interface modelling.

\color{black}

We have avoided exploring a wide variety of battery electrode/electrolyte
materials and battery concepts such as lithium-sulfur (Li-S), lithium-oxygen
(Li-O$_2$), and conversion cathode materials on purpose.  The latter areas
have been subjects of modelling overviews, while a critical review of modelling
methods has been lacking.  The unifying modelling concepts discussed in this
work focus on metallic conductor electrodes and reasonably well-characterized
surface films or solid electrolytes.  These should be broadly applicable to
all battery electrodees like Na anodes, lithiated graphite, and lithiated
silicon.  One exception is multivalent (e.g., Mg) batteries, where
surface films have been avoided to prevent slow multivalent ion transport.
Li-S battery apply Li-metal cathodes and often carbon-based cathodes that
trap sulfur; these electrodes are metallic and contain surface CEI films
too.  Carbon cathode/Li$_2$O$_2$ interfaces in Li-air batteries also fit this
category.  All-solid state with metallic anodes would be prototype systems
to apply these computational ideas, but the thicker solid electrolyte 
compared to surface films in liquid electrolyte batteries may require special
treatment (e.g., those devised for DFT modelling of Mott-Schottky contacts)
address the larger space charge region.  
Computing overpotential effects on kinetics in kinetics-limited conversion
cathode materials may be a particularly fruitful area to apply our approach;
the calculations of reaction barriers at non-overpotential conditions require
careful tuning of the electronic voltage at the interface.
Interfaces between transition metal oxide cathodes and liquid electrolytes may
present the most significant challenges.  While the non-metallic nature in
many such oxides can be circumvented with by adding a computational ``current
collector'' at the bottom, the very thin surface film (``CEI'') covering oxide
surfaces, which has mostly unknown atomic structures, hinders assignment of
voltages and predictions of degradation and transport reaction kinetics.
In the final analysis, at this and all other battery
interfaces, the voltage is an imposed, DFT-constructed  quantify;
the question is whether model interfacial structures that
provide the imposed voltages yield electric fields that reflect
experimental conditions.

\color{black}

\section*{Acknowledgement}
 
We thank Mira Todorova, Minoru Otani, Axel Gross, Harald Oberhofer, and De-en
Jiang for discussions during the 2019 Telluride Workshop on Electrochemical
Interfaces.  We also thank Quinn Campbell and Catalin Spataru for
commenting on the manuscript.  Above all, we gratefully acknowledge 
Michiel Sprik's pioneering research.

The Li/LiOH work is funded by Nanostructures for Electrical Energy Storage
(NEES), an Energy Frontier Research Center funded by the U.S. Department
of Energy, Office of Science, Office of Basic Energy Sciences under Award
Number DESC0001160.  The Li/LiF work is funded by the Laboratory Directed
Research and Development Program at Sandia National Laboratories.  The
Al/Al$_2$O$_3$ work is funded by the Advanced Strategic Computing (ASC) Program.
Sandia National Laboratories is a multi-mission laboratory managed and operated
by National Technology and Engineering Solutions of Sandia, LLC, a wholly owned
subsidiary of Honeywell International, Inc., for the U.S. Department of
Energy’s National Nuclear Security Administration under contract DE-NA0003525.
This paper describes objective technical results and analysis.  Any subjective
views or opinions that might be expressed in the document do not necessarily
represent the views of the U.S. Department of Energy or the United States
Government.




\begin{references}
\bibitem{stability}
Here ``electrochemical stability'' is defined as the empirically-determined
voltage range over which interfaces are stable.  It involves a mixture
of thermodynamics, kinetics, and electron transfer considerations.\cite{batt}
\bibitem{note1}
In contrast, multivalent batteries with M$^{2+}$ and M$^{3+}$ charge
carriers must use electrolyte that does not form surface films, because 
these ions moves slowly through solid films.  These battery concepts
requires substantial development.
\end{references}
\end{document}